\documentclass[aps,reprint,twocolumn,amsmath,amssymb,showpacs]{revtex4-1}
\usepackage[hidelinks]{hyperref}
\usepackage{graphicx,color,bm}

\begin{document}

\title{Negative refraction in Weyl semimetals}

\author{M.~Shoufie Ukhtary} 
\email{shoufie@flex.phys.tohoku.ac.jp}
\author{Ahmad R. T. Nugraha}
\author{Riichiro Saito}

\affiliation{Department of Physics, Tohoku University, Sendai
  980-8578, Japan}

%%%%%%%%%%%%%%%%%%%%%%%%%%%%%%%%%%%%%%%%%%%%%%%%%%%%%%%%%%%%%%%%%%%%%%%%%%%%%%%%
%                                  Abstract
%%%%%%%%%%%%%%%%%%%%%%%%%%%%%%%%%%%%%%%%%%%%%%%%%%%%%%%%%%%%%%%%%%%%%%%%%%%%%%%%

\begin{abstract}
  We theoretically propose that Weyl semimetals may exhibit negative
  refraction at some frequencies close to the plasmon frequency,
  allowing transverse magnetic (TM) electromagnetic waves with
  frequencies smaller than the plasmon frequency to propagate in the
  Weyl semimetals.  The idea is justified by the calculation of
  reflection spectra, in which \textit{negative} refractive index at
  such frequencies gives physically correct spectra.  In this case, a
  TM electromagnetic wave incident to the surface of the Weyl
  semimetal will be bent with a negative angle of refraction.  We
  argue that the negative refractive index at the specified
  frequencies of the electromagnetic wave is required to conserve the
  energy of the wave, in which the incident energy should propagate
  away from the point of incidence.
 \end{abstract}

\date{\today}
\maketitle

%%%%%%%%%%%%%%%%%%%%%%%%%%%%%%%%%%%%%%%%%%%%%%%%%%%%%%%%%%%%%%%%%%%%%%%%%%%%%%%%
%                               Introduction
%%%%%%%%%%%%%%%%%%%%%%%%%%%%%%%%%%%%%%%%%%%%%%%%%%%%%%%%%%%%%%%%%%%%%%%%%%%%%%%%
\section{Introduction}

Negative refraction phenomenon has attracted many interests since its
prediction by Veselago a half century
ago.~\cite{veselago1968electrodynamics} Veselago predicted that if a
material possesses simultaneous negative dielectric constant
($\varepsilon$) and magnetic permeability ($\mu$), it will give a
negative refractive index. The negative refractive index will lead to
some unusual properties of the light, such as negative refraction and
reversed Doppler and Cherenkov
effects.~\cite{veselago1968electrodynamics,cubukcu2003electromagnetic,pendry2000negative,habe2015spin,cheianov2007focusing,smith2004metamaterials}
By utilizing negative refraction, in which the light will be bent in
an unusual way with an angle of refraction negative to the normal
direction of the material surface, one may be able to construct a
superlens whose resolution is smaller than the light wave
length.~\cite{cubukcu2003electromagnetic,pendry2000negative,smith2004metamaterials}
A better Cherenkov radiation detector can also be realized based on
the material having a negative refractive index, which is useful in
the field of accelerator
physics.~\cite{lu2003vcerenkov,ziemkiewicz2015cherenkov} However,
materials having simultaneous negative $\varepsilon$ and $\mu$ have
not been found in nature so far.

To realize negative refraction, many researchers developed artificial
structures that are called as
metamaterials.~\cite{shalaev2007optical,boltasseva2008fabrication,padilla2006negative,valentine2008three}
These structures usually contain an array of split ring
resonators~\cite{smith2004metamaterials,ishikawa2005negative,moser2005terahertz,bilotti2007design}
or dielectric photonic crystals with periodically modulated
$\varepsilon$ and
$\mu$,~\cite{cubukcu2003electromagnetic,smith2004metamaterials,parimi2003photonic}
which are often complicated to fabricate. To overcome the
difficulties, in this paper we predict that negative refraction can
take place in a bulk Weyl semimetal (WSM) even without having negative
$\mu$ and without constructing complicated structure.  The WSM is a
three-dimensional material having a pair of Dirac cones separated in
the $k$ space in its energy dispersion shown in
Fig.~\ref{fig1}(a).~\cite{hofmann2016surface,burkov2011weyl,vazifeh2013electromagnetic,koshino2016magnetic,ominato2015quantum}
An example of the WSM is pyrochlore
($\mathrm{Eu_2Ir_2O_7}$).~\cite{hofmann2016surface,sushkov2015optical}
In each cone, the valence and conduction bands coincide at the
so-called Weyl nodes. The presence of a pair of separated Dirac cones
is the consequence of symmetry breaking in the WSM, which induces the
Hall current, even without magnetic
field.~\cite{hofmann2016surface,burkov2011weyl,vazifeh2013electromagnetic}
This phenomenon is known as the anomalous Hall effect, which is
responsible for the tensor form of the dielectric function of the
WSM~\cite{zyuzin2015chiral,hofmann2016surface}. In this work, we
predict that the EM wave can propagate through WSM even though the
frequency is smaller than plasmon frequency. This propagation requires
the refractive index of WSM to be negative in order to conserve the
energy, that will be shown in this paper.

%%%%%%%%%%%%%%%%%%%%%%%%%%%%%%%%%%%%%%%%%%%%%%%%%%%%%%%%%%%%%%%%%%%%%%%%%%%%%%%%
%                            Model and Methods
%%%%%%%%%%%%%%%%%%%%%%%%%%%%%%%%%%%%%%%%%%%%%%%%%%%%%%%%%%%%%%%%%%%%%%%%%%%%%%%%
\section{Model and Methods}
\begin{figure}[t]
  \centering\includegraphics[width=85mm]{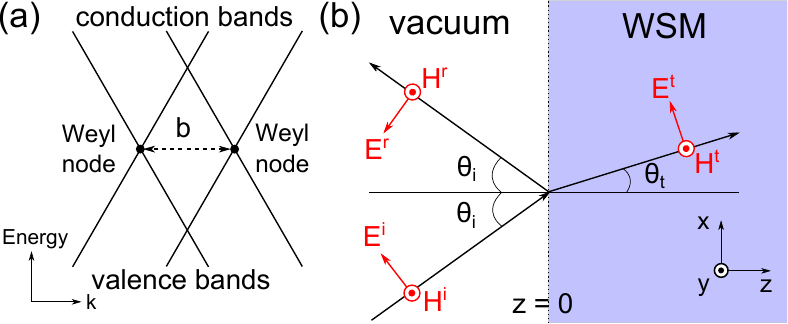}
  \caption{(a) Schematic of energy dispersion of WSM showing a pair of
    Dirac cones with two Weyl nodes represented by dots, separated by
    the wave vector $b$. (b) A TM wave coming to $xy$ surface of WSM
    at angle $\theta_i$ and transmitted to WSM at angle
    $\theta_t$.}
\label{fig1}
\end{figure}

The electromagnetic response of WSM can be derived from the formula of
action for the electromagnetic
field.~\cite{zyuzin2015chiral,vazifeh2013electromagnetic,grushin2012consequences}
Here, we will give brief derivation of the electromagnetic response of
WSM represented by electric displacement vector $\textbf{D}$. The more
detailed derivation is given by Zyuzin and
Burkov~\cite{zyuzin2012topological,zyuzin2012weyl} or Hosur and
Qi.~\cite{hosur2013recent} The action of electromagnetic field is
given by,
\begin{equation}
S_{\theta}=-\frac{e^2}{8\pi^2\hbar}\int dt d\textbf{r}\partial_\gamma\theta\epsilon^{\gamma\nu\rho\eta}A_\nu\partial_\rho A_\eta,
\label{eq:action}
\end{equation}
where $A_\nu$ is electromagnetic potential,
$\epsilon^{\gamma\nu\rho\eta}$ is the Levi-Civita tensor and
 each index $\gamma,\nu,\rho,\eta$ takes values $0,1,2,3$. The term $\theta$ is called the axion angle given by
$\theta=2(\textbf{b}\cdot\textbf{r})$, where $\textbf{b}$ is a wave
vector separating the Weyl nodes [see Figure~\ref{fig1}(a)]. The
current density $j_\nu$ is given by varying the action with respect to
electromagnetic potential,
\begin{align}
j_\nu\equiv\frac{\delta S_\theta}{\delta
  A_\nu}=\frac{e^2}{4\pi^2\hbar}\partial_\gamma\theta\epsilon^{\gamma\nu\rho\eta}
\partial_\rho A_\eta.
\label{eq:current}
\end{align}
By writing $\textbf{E}=-(\nabla A_0)-\partial_0\textbf{A}$, Eq.~(\ref{eq:current}) gives the Hall current $\textbf{j}=\frac{e^2}{4\pi^2\hbar}\nabla\theta\times
\textbf{E}$, which gives
additional terms in $\textbf{D}$ of the normal metals as the second
term of Eq.~(\ref{eq:displacemet}). We can write the electric
displacement vector as follows,
\begin{equation} 
\textbf{D}=\varepsilon_0\varepsilon_b\left(1-\frac{\omega_p^2}{\omega^2}\right)\textbf{E}+\frac{ie^2}{4\pi^2\hbar\omega}(\nabla\theta)\times\textbf{E},
\label{eq:displacemet}
\end{equation}
where $\omega_p$ is the plasmon frequency, $\varepsilon_b$ is the
background dielectric constant. Hereafter, we consider a particular
value of the dielectric constant, $\varepsilon_b=13$, which was
measured in pyrochlore.~\cite{hofmann2016surface,sushkov2015optical}
The first term of Eq.~(\ref{eq:displacemet}) is the Drude dielectric
function, which is similar to normal metals (NMs). The appearance of
Hall current without external magnetic field is known as anomalous
Hall effect given by the second term of
Eq.~(\ref{eq:displacemet}). The anomalous Hall current only depends on
the structure of the electron dispersion of WSM represented by
$\theta$. Due to the anomalous Hall effect, the dielectric tensor has
non-zero off-diagonal terms, which can be written as
\begin{equation}
  \varepsilon
  =
\begin{bmatrix}
  \varepsilon_1
  & 0 &i\varepsilon_2\\
  0 & \varepsilon_1 & 0\\
  -i\varepsilon_2 & 0 & \varepsilon_1
\end{bmatrix}
\label{eq:tensor}
\end{equation}
where we assume that $\textbf{b}$ lies in the direction of $y$,
$\textbf{b}=b\mathbf{\hat{y}}$, and that $\varepsilon_1$ and
$\varepsilon_2$ are expressed by
\begin{align}
  \varepsilon_1 &= \varepsilon_0\varepsilon_b
                  \left(1-\frac{1}{\Omega^2}\right)\label{eq:e1},\\
  \varepsilon_2 &= \varepsilon_0\varepsilon_b
                  \left(\frac{\Omega_{b}}{\Omega}\right)\label{eq:e2},
\end{align}
with $\Omega=\omega/\omega_p$ and
$\Omega_{b}=e^2b/(2\pi^2\varepsilon_0\varepsilon_b\hbar\omega_p)$ as
dimensionless quantities.  We take $\Omega_{b}=0.5$ as a fixed
parameter throughout this paper, otherwise it will be mentioned.
Similar to NMs, in the WSM we have $\varepsilon_1>0~(\varepsilon
_1<0)$ if $\Omega>1~(\Omega <1)$.

In order to calculate the reflection and transmission spectra of a
bulk WSM, we will determine the refractive index of the WSM
$(n_w)$. Suppose that we have a transverse magnetic (TM) wave incident
at angle $\theta_i$ from vacuum to a WSM as shown in
Fig.~\ref{fig1}(b) where $E^{i}$, $E^{r}$ and $E^{t}$ ( $H^{i}$,
$H^{r}$ and $H^{t}$) are the incident, reflected and transmitted
electric (magnetic) fields, respectively. The transmitted wave
propagates toward positive $z$ direction inside WSM, while the reflected wave
propagates toward negative $z$ direction.  Due to the vanishing
$\varepsilon_{xy}$ and $\varepsilon_{zy}$, the direction of electric
field inside WSM does not rotate.  By using Eq.~\eqref{eq:tensor}, we
can write down the equation $\textbf{D}=\mathbf{\hat{\varepsilon}}
\textbf{E}$ for the TM wave inside the WSM as follows,
\begin{equation}
  \begin{bmatrix}
    D^{t}_x\\D^{t}_y\\D^{t}_z
  \end{bmatrix}
  =
  \begin{bmatrix}
    \varepsilon_1 & 0 & i\varepsilon_2\\
    0 & \varepsilon_1 & 0\\
    -i\varepsilon_2 & 0 & \varepsilon_1
  \end{bmatrix}
  \begin{bmatrix}
    E^{t}_x \\ 0 \\ E^{t}_z
  \end{bmatrix}
  \label{eq:D}
\end{equation}
where $D^{t}$ and $E^{t}$ are the displacement and electric fields
inside the WSM.  From Maxwell's equations, we get a differential
equation for the EM wave as follows;
\begin{equation}
  \nabla\times\nabla\times\textbf{E}^t
  =-\nabla^2\textbf{E}^t+\nabla
  \left(\nabla\cdot\textbf{E}^t\right)=\omega^2\mu_0\textbf{D}^t.
  \label{eq:waveeq}
\end{equation}
Since the solutions of $\textbf{E}^t$ and $\textbf{D}^t$ are
proportional to $\exp\left[i\omega
  n_w/c~\left(\textbf{s}\cdot\textbf{r}\right)\right]$, where
$\textbf{s}=\left(\sin\theta_t,0,\cos\theta_t\right)$ is the unit wave
vector, we can obtain from Eq.~(\ref{eq:waveeq}),
\begin{equation}
  \frac{1}{\mu_0}\left(\frac{n_w}{c}\right)
  \left[\textbf{E}^t-\textbf{s}
    \left(\textbf{s}\cdot\textbf{E}^t\right)\right]=\textbf{D}^t.
  \label{eq:waveeq2}
\end{equation}
From Eqs.~(\ref{eq:D}) and~(\ref{eq:waveeq2}), we get the following relations,
\begin{equation}
  E^t_x = \frac{\varepsilon_1D^t_x -
  i\varepsilon_2D^t_z}{\varepsilon_1^2 - \varepsilon_2^2}
  ,~\textrm{and}\quad E^t_z = \frac{i\varepsilon_2D^t_x +
  \varepsilon_1D^t_z}{\varepsilon_1^2 - \varepsilon_2^2}.
\label{eq:E}
\end{equation}
Inserting Eq.~(\ref{eq:E}) to Eq.~(\ref{eq:waveeq2}), we obtain
simultaneous equations of $E^t_x$ and $E^t_z$ as follows:
\begin{widetext}
\begin{equation}
\label{eq:det}
\begin{bmatrix}
  (1-s_x^2)\varepsilon_1-is_xs_z\varepsilon_2-
  \mu_0\left(\frac{c}{n_w}\right)^2\left(\varepsilon_1^2-\varepsilon_2^2\right)
  & -i\left(1-s_x^2\right)\varepsilon_2-s_xs_z\varepsilon_1
  \\ i\left(1-s_z^2\right)\varepsilon_2-s_xs_z\varepsilon_1 &
  (1-s_z^2)\varepsilon_1+is_xs_z\varepsilon_2-\mu_0
  \left(\frac{c}{n_w}\right)^2\left(\varepsilon_1^2-\varepsilon_2^2\right)
\end{bmatrix}
\begin{bmatrix}
  E^t_x\\E^t_z
\end{bmatrix}
=0
\end{equation}
\end{widetext}
In order to have nontrivial solutions of $\textbf{E}^t$, the determinant of the
$2\times 2$ matrix in Eq.~(\ref{eq:det}) should vanish:
\begin{equation}
\frac{\mu_0c^2\left(\varepsilon_1^2-\varepsilon_2^2\right)}{n_w^4}
\left[-n_w^2\varepsilon_1+c^2\mu_0
  \left(\varepsilon_1^2-\varepsilon_2^2\right)\right]=0.
\label{eq:det2}
\end{equation}
from which, we obtain $n_w$,
\begin{align}
  n_w = \pm
  c\sqrt{\mu_0\left(\varepsilon_1^2-\varepsilon_2^2\right)/\varepsilon_1}
  \equiv n_w^{\pm},
\label{eq:nw}
\end{align}
where the $n_w^{+}$
($n_w^{-}$)
solution corresponds to the positive (negative) wave vector inside
the WSM. If we put $\varepsilon_2=0$
in Eq.~(\ref{eq:nw}), we can obtain the refractive index of NM.

\begin{figure}[h]
 \includegraphics[width=85mm]{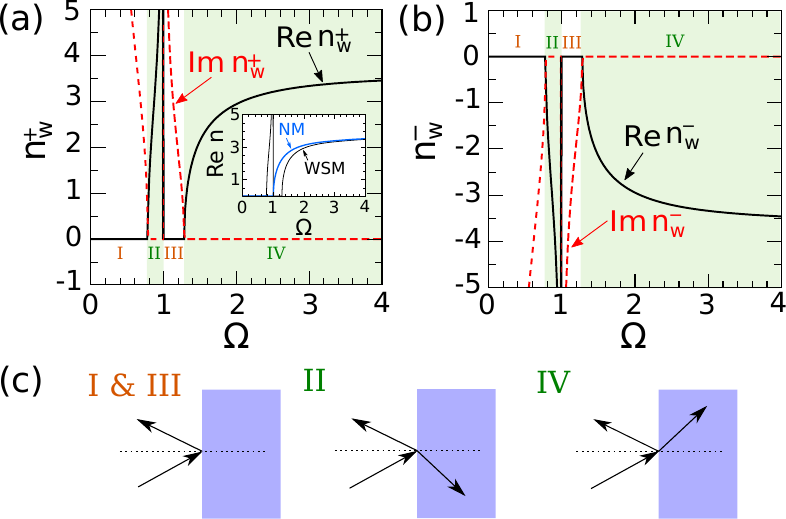}
 \caption{ (a) The refractive index of WSM for TM wave $(n_{w})$ as a
   function of $\Omega$ for the positive solution of Eq.~(\ref{eq:nw})
   ($n_w^{+}$).  Solid and dashed lines are the real and imaginary
   parts of $n_{w}^{+}$, respectively. We use $\Omega_b=0.5$ for the
   WSM.  The plot is divided into four regions.  Inset: The real part
   of refractive index $(n)$ for the WSM compared with a normal metal
   (NM). (b) The refractive index of WSM for TM wave $(n_{w})$ as a
   function of $\Omega$ for the negative solution of Eq.~(\ref{eq:nw})
   ($n_w^{-}$). (c) Schematics of EM wave propagations to the WSM for
   all the four regions of panel (a) and (b).}
\label{fig2}
\end{figure} 

%%%%%%%%%%%%%%%%%%%%%%%%%%%%%%%%%%%%%%%%%%%%%%%%%%%%%%%%%%%%%%%%%%%%%%%%%%%%%%%%
%                            Model and Methods
%%%%%%%%%%%%%%%%%%%%%%%%%%%%%%%%%%%%%%%%%%%%%%%%%%%%%%%%%%%%%%%%%%%%%%%%%%%%%%%%
In Fig.~\ref{fig2}(a) and (b) we plot $n_{w}$ as a function of
$\Omega$ for the positive solution of Eq.~(\ref{eq:nw})
[Fig.~\ref{fig2}(a)] and the negative solution of Eq.~(\ref{eq:nw})
[Fig.~\ref{fig2}(b)].  The solid and the dashed lines correspond to
the real and imaginary parts of $n_w$, respectively. It is noted that
$n_w^{\pm}$ at each frequency is either purely real or purely
imaginary, because we neglect the effects of the impurity and
scattering of charge in Eq.~(\ref{eq:e1}).  Therefore, the wave vector
$\omega n_w^{\pm}/c$ can be either real or imaginary depending on
$n_w^{\pm}$.  The real (imaginary) wave vector represents a
propagating (decaying) wave.

Here we divide our results into four regions as shown in
Fig.~\ref{fig2}(c): region I $\left(0~\leq~\Omega~\leq
\Omega_-\right)$, region II $\left(\Omega_-~\leq~\Omega~\leq
1\right)$, region III $\left(~1\leq~\Omega~\leq \Omega_+\right) $ and
region IV $\left(~\Omega_+\leq\Omega\right) $, where $\Omega_{\pm}$
are frequencies that give $n_w=0$ [Eqs.~(\ref{eq:e1}), (\ref{eq:e2}),
  (\ref{eq:nw})].
\begin{align}
  \Omega_{\pm}=1/2\left(\pm\Omega_{b}+\sqrt{4+\Omega_{b}^2}\right).
\label{eq:opm}
\end{align}
 As defined before,
 $\Omega_{b}=e^2b/(2\pi^2\varepsilon_0\varepsilon_b\hbar\omega_p)$,
 where $b=0.37\textrm{\AA}^{-1}$ for pyrochlore~\cite{sushkov2015optical}
 and plasmon frequency is given by~\cite{hofmann2016surface}
	\begin{align}
  \omega_p=\sqrt{\frac{4\alpha}{3\pi}}\frac{E_{\textrm{F}}}{\hbar}
\label{eq:wp}
\end{align}
with $\alpha=\frac{e^2}{\hbar v_{\textrm{F}}\varepsilon_0\varepsilon_b}$ and
$v_{\textrm{F}}=4\times
10^{7}~\textrm{cm/s}$.~\cite{sushkov2015optical}

From Fig.~\ref{fig2}(a), it is important to point out that we may have
a propagating wave even at frequencies smaller than plasmon frequency
$(\Omega< 1)$, in the shaded region II, which is in contrast with NM
where an EM wave can propagate if $\Omega> 1$ [see inset of
  Fig.~\ref{fig2}(a)]. As shown in the inset of Fig.~\ref{fig2}(a),
the refractive indices of WSM and NM differ only near $\Omega\simeq
1$.  At $\Omega\gg 1$, they both converge to the value of
$n\approx\sqrt{\varepsilon_b}$. It is important to note that the
negative solution of Eq.~(\ref{eq:nw}) ($n_w^{-}$) is assigned to have
propagating wave toward positive $z$ direction in the region II, which
will be shown later.

Let us calculate the reflection and transmission spectra. In NM with
applied external magnetic field, the polarization of EM wave undergoes
rotation as it enters the material if the direction of propagation is
parallel to the direction of applied external magnetic field making
the wave polarization not linear. In our case of WSM, we choose the propagation
direction of \textit{the purely TM wave} $(E_y=0)$ to be
\textit{perpendicular} to the "effective applied magnetic field", which is in
the direction of the $\textbf{b}=b\mathbf{\hat{y}}$. Therefore, we
expect no rotation of polarization and the wave polarization keeps
linear as TM wave. This fact can also be deduced from the vanishing
$\varepsilon_{xy}$ and $\varepsilon_{zy}$. As shown in
Fig.~\ref{fig1}(b), the incident, reflected, and transmitted electric
fields $\textbf{E}_{i}$, $\textbf{E}_{r}$ and $\textbf{E}_{t}$ can be
written as
\begin{align}
  \textbf{E}^{i}(z) &= \left(\cos\theta_i, 0,
-\sin\theta_i\right)E^{i}_{0}\exp(ik_{vz}z),\label{eq:efield1}\\
  \textbf{E}^{r}(z) &= \left(-\cos\theta_i, 0,
-\sin\theta_i\right)E^{r}_{0}\exp(-ik_{vz}z),\label{eq:efield2}\\
  \textbf{E}^{t}(z) &= \left(\cos\theta_t, 0,
-\sin\theta_t\right)E^{t}_{0}\exp(ik_{wz}^{\pm}z),\label{eq:efield}
\end{align}
with $k_{vz}=(\omega/c)~\cos\theta_i$ and $k_{wz}^{\pm}=\omega
(n_{w}^{\pm}/c)~\cos\theta_t$. The angles $\theta_i$ and $\theta_t$
are related each other by the Snell's law
$\sin\theta_i=n_w^{\pm}\sin\theta_t$. The magnetic fields in the $y$
direction can be obtained from the relations $H^{i,r}_y=i\omega\int
\varepsilon_0 E^{i,r}_x dz$ and $H^{t}_y=i\omega\int D^{t}_x dz$,
where $D^{t}_x=\varepsilon_1E^{t}_x+i\varepsilon_2E^{t}_z$ is obtained
from Eq.~(\ref{eq:D}). Then, the magnetic fields can be written as
\begin{align}
  \textbf{H}^{i}(z) =& \frac{\omega\varepsilon_0}{k_{vz}}
  \left(0,\cos\theta_i,0\right)
  E^{i}_{0}\exp(ik_{vz}z),\label{eq:hfield1}\\ 
  \textbf{H}^{r}(z) =&
  \frac{\omega\varepsilon_0}{k_{vz}}
  \left(0,\cos\theta_i,0\right)
  E^{r}_{0}\exp(-ik_{vz}z),\label{eq:hfield2}\\
  \textbf{H}^{t}(z)=&\frac{\omega}{k_{wz}^{\pm}}
  \left(0,\varepsilon_1\cos\theta_t-i\varepsilon_2\sin\theta_t,0\right)
  E^{t}_{0}\nonumber\\&\times \exp(ik_{wz}^{\pm}z).
\label{eq:hfield}
\end{align} 

After defining the EM fields in both media, we can write down boundary
conditions of the EM wave at incidence surface $\left(z=0\right)$ as
follows,
\begin{align}
  &E^{i}_{0}\cos\theta_i-E^{r}_{0}\cos\theta_i =
  E^{t}_{0}\cos\theta_t,\label{eq:bond1}
\end{align}
and
\begin{align}
 &\frac{\omega\varepsilon_0}{k_{vz}} \left(E^{i}_{0} 
  \cos\theta_i+E^{r}_{0}\cos\theta_i\right) \notag \\
  &= \frac{\omega}{k_{wz}^{\pm}}\left(\varepsilon_1E^{t}_{0}
  \cos\theta_t-i\varepsilon_2E^{t}_{0}\sin\theta_t\right)\label{eq:bond2},
\end{align}
where Eqs.~(\ref{eq:bond1}) and (\ref{eq:bond2}) describe the
continuity for the tangential components of electric fields and
magnetic fields at $z=0$, respectively. Reflection coefficient
$r=E^{r}_{0}/E^{i}_{0}$ and transmission coefficient
$t=E^{t}_{0}/E^{i}_{0}$ are given by
\begin{align}
r&=1-t  \frac{\cos\theta_t}{\cos\theta_i},
\label{eq:rp}
\end{align}
and
\begin{align}
  t&=\frac{2k_{wz}^{\pm}\varepsilon_0\cos\theta_i}{k_{vz}
    \left(\varepsilon_1\cos\theta_t -
    i\varepsilon_2\sin\theta_t\right) +
    k_{wz}^{\pm}\varepsilon_0\cos\theta_t}.
\label{eq:tp}
\end{align}  

\section{Results and Discussion}

\begin{figure}[h!]
 \includegraphics[width=85mm]{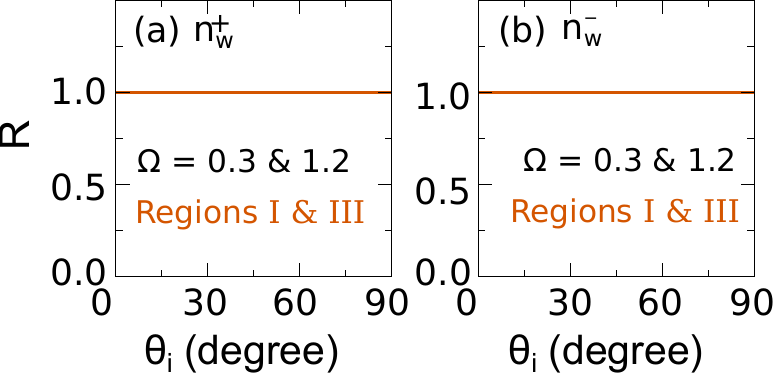}
 \caption{ The reflection probability (R) of the EM wave in a WSM for
   region I [$\Omega=0.3$] and region III [$\Omega=1.2$] for (a)
   $n_w^{+}$ and (b) $n_w^{-}$.}
\label{fig3}
\end{figure}

In Fig.~\ref{fig3}, we plot the reflection probability defined by
$R=\left|r\right|^2$ as a function of $\theta_i$ for region I and III
(that is $\Omega=0.3$ and 1.2, respectively). Fig.~\ref{fig3}(a)
shows $R$ if we use $n_w^{+}$ and Fig.~\ref{fig3}(b) shows $R$ if we
use $n_w^{-}$. From Fig.~\ref{fig3}, we can see that the incident EM
wave will be totally reflected $R=1$ for all $\theta_i$ for both
$n_w^{\pm}$ as shown in Fig.~\ref{fig3}(a) and (d), due to the purely
imaginary $n_w^{\pm}$ given in Fig.~\ref{fig2}(a).  The
$\textbf{E}^{t}$ is decaying inside WSM, hence no transmitted energy
into WSM. The most interesting case is region II, where we predict
that WSM acquires a negative refractive index. In region II, we have a
real $n_w^{\pm}$, which means that the wave propagation inside WSM is
allowed, even though the wave frequency is smaller than the plasmon
frequency.

Normally, we use $n_w^{+}$ which gives a positive value of
$k_{wz}^{\pm}$ because the transmitted wave propagates toward positive $z$
direction [see Eq.~(\ref{eq:efield})].  However, $n_w^{+}$ for the
transmitted wave in region II gives an unphysical $R>1$ , which means
that at the point of incidence there is a flux of energy coming from
the WSM side. We can infer from Eqs.~(\ref{eq:rp}) and (\ref{eq:tp})
that $R>1$ if $n_w^{+}$ is selected for region II.  The reflection
coefficient $r$ can be written as
\begin{align}
 r&=\frac{A-B-iC}{A+B-iC}\\
	&=r_{1}+ir_{2}
\label{eq:rr}
\end{align}  
where $A=\varepsilon_1\cos\theta_i\cos\theta_t$, $B=
n_w^{\pm}\varepsilon_0\cos^2\theta_t$,
$C=\varepsilon_2\cos\theta_i\sin\theta_i$. The reflection probability
can be obtained from $R=r_{1}^2+r_{2}^2$, where we define
\begin{align}
 r_{1}&=\frac{(A+B)(A-B)+C^2}{(A+B)^2+C^2}\label{eq:rria}\\
r_{2}&=\frac{2CB}{(A+B)^2+C^2}.
\label{eq:rrib}
\end{align}
$R>1$ if either $r_{1}>1$ or $r_{2}>1$. Let us investigate the case of
$r_1$. From Eq.~(\ref{eq:rria}), we can define the requirement in
order to have $r_1<1$ giving us physically sound $R<1$, otherwise we will have unphysical $R>1$,
\begin{align}
\left|A-B\right|&<\left|A+B\right|\\ &\textrm{or}\nonumber\\ \left|\varepsilon_1\cos\theta_i-
n_w^{\pm}\varepsilon_0\cos\theta_t\right|&<\left|\varepsilon_1\cos\theta_i+
n_w^{\pm}\varepsilon_0\cos\theta_t\right|.
\label{eq:req}
\end{align}

\begin{figure}[h!]
 \includegraphics[width=70mm]{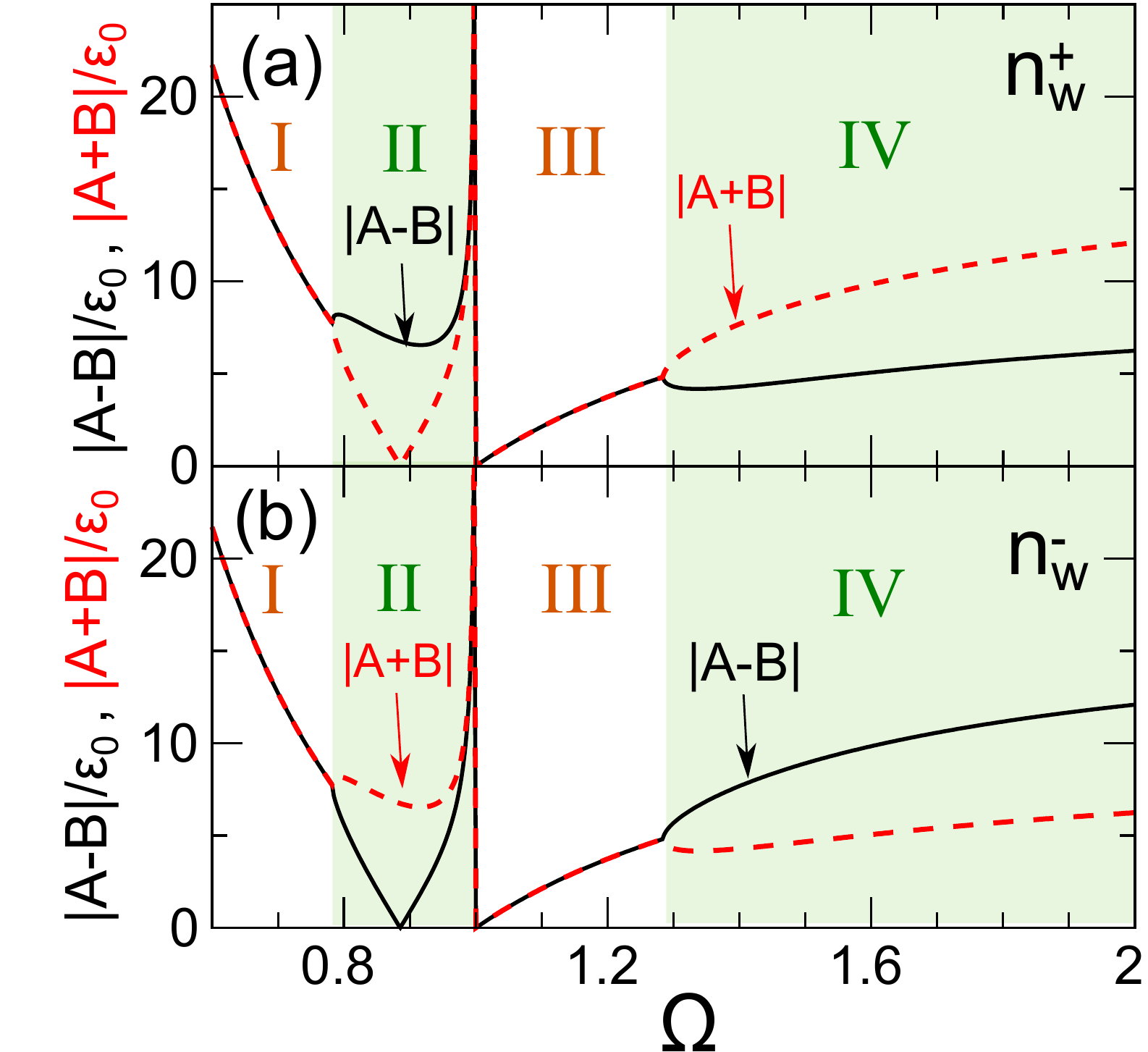}
 \caption{$\left|A-B\right|$ and $\left|A+B\right|$ as a function of
   $\Omega$ if we use (a) $n_w^{+}$ and (b) $n_w^{-}$. In region II,
   $n_w^{-}$ is selected to fulfill Eq.~(\ref{eq:req}), while in
   region IV, $n_w^{+}$ is selected. Otherwise, we will have
   unphysical $R>1$ in both region. }
\label{fig4}
\end{figure}

To better visualize Eq.~(\ref{eq:req}), we plot $\left|A-B\right|$ and
$\left|A+B\right|$ as a function of $\Omega$. From Fig.~\ref{fig4}(a),
where $n_w^{+}$ is selected, $\left|A-B\right|>\left|A+B\right|$ in
region II, which does not fulfill Eq.~(\ref{eq:req}) giving the
unphysical $R>1$. On the other hand, from Fig.~\ref{fig4}(b), where
$n_w^{-}$ is selected, $\left|A-B\right|<\left|A+B\right|$ in region
II, which fulfills Eq.~(\ref{eq:req}) and we can have physically
correct $R<1$. This negative solution ($n_w^{-}$) should be selected
only for region II, because if we apply $n_w^{-}$ to region IV, we
have an unphysical $R>1$, which is shown by Fig.~\ref{fig4}(b), in
which $\left|A-B\right|>\left|A+B\right|$ for region IV. We argue
later that the reason why $n_w^{-}$ is selected in region II for having transmitted wave
toward positive $z$-direction, is due to the energy conservation.

The negative refractive index of WSM in region II will cause the wave
refracted negatively, which means that the refracted angle $\theta_t$
is negative.  The refractive index also means that the wave vector of
transmitted wave ($k_{wz}^{-}$) is
negative.~\cite{saleh1991fundamentals,smith2000negative,ramakrishna2005physics,woodley2006backward}
The negative wave vector does not mean that the transmitted wave
propagates backward, which violates the conservation of energy. The
direction of propagation is better determined by the direction of the
Poynting vector. By using Eqs.~(\ref{eq:efield}) and (\ref{eq:hfield})
at $z=0$, the power per unit cross section transmitted in the
direction of $z$ can be expressed as
\begin{align}
  I_t &=\textbf{S}_t\cdot\hat{\textbf{z}}\nonumber\\
	&=\frac{1}{2}\textrm{Re}\left[\textbf{E}^{t}(0)
       \times\textbf{H}^{\textbf{*}t} (0)\right]\cdot \hat{\textbf{z}}\notag\\
     &=\frac{c\left|t\right|^2\left| E^{i}_{0}\right|^2}{2n_w^{\pm}}
       \varepsilon_1\cos\theta_t,
       \label{eq:intensity}
\end{align}

\begin{figure}[t]
  \includegraphics[width=60mm]{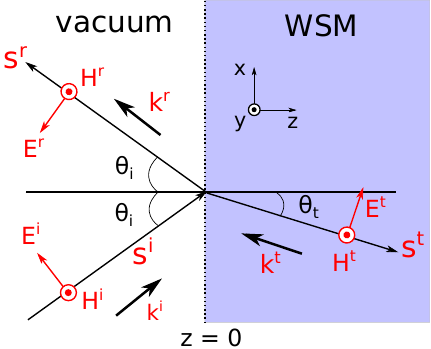}
  \caption{The negative refraction in WSM. $\textbf{S}_t$ is the
    transmitted Poynting vector. $\textbf{k}_i,　\textbf{k}_r,
    \textbf{k}_t$ are incident, reflected and transmitted wave
    vectors, respectively.}
\label{fig5}
\end{figure}

In order to have transmitted power propagate toward positive $z$
direction, Eq.~(\ref{eq:intensity}) should have a positive
value. Since $\varepsilon_1<0$ in region II [Eq.~(\ref{eq:e1})], while
$\left|t\right|^2$, $\left| E^{i}_{0}\right|^2$, and $\cos\theta_t
>0$, $n_w^{\pm}$ has to be \textit{negative} $(n_w^{-})$ in order to
have $I_t>0$. On the other hand, $n_w^{+}$ is selected in region IV,
because $\varepsilon_1>0$. We refer the transmitted wave as backward
wave because the transmitted wave vector points towards negative $z$-direction shown by Fig.~\ref{fig5}, otherwise it is forward wave. In short, the negative refraction is needed
for the propagation of the EM wave with frequency smaller than the
plasmon frequency to \textit{conserve energy}.

To show the negative refraction more explicitly, we calculate the
tangential component of the transmitted Poynting vector with respect
to the interface. The tangential component of Poynting vector is given
by,
\begin{align}
  \textbf{S}_t\cdot\hat{\textbf{x}}=&=\frac{1}{2}\textrm{Re}\left[\textbf{E}^{t}(0)
       \times\textbf{H}^{\textbf{*}t} (0)\right]\cdot \hat{\textbf{x}}\nonumber\\
     &=\frac{c\left|t\right|^2\left| E^{i}_{0}\right|^2}{2(n_w^{\pm})^2}
       \varepsilon_1\sin\theta_i.
       \label{eq:sx}
\end{align}
Because at region II, $\varepsilon_1<0$ and all other terms are
positive, then $\textbf{S}\cdot\hat{\textbf{x}}<0$, which means that
we have negative refraction. Therefore, at region II, we expect the
light is transmitted as backward wave with negative refraction shown
by Fig.~\ref{fig5}.

It is also interesting to compare our case with hyperbolic
metamaterial. The negative refraction phenomenon in WSM is similar to
hyperbolic metamaterials, where we can obtain negative refraction
without having negative magnetic permeability. In hyperbolic
metamaterials, due to the anisotropy of its dielectric tensor with
respect to crystal axis, where the parallel and perpendicular
component of dielectric tensor are opposite sign
$(\bar{\bar{\varepsilon}}=\varepsilon_\perp\hat{\textbf{x}}\hat{\textbf{x}}+\varepsilon_\parallel[\hat{\textbf{y}}\hat{\textbf{y}}+\hat{\textbf{z}}\hat{\textbf{z}}]$,
with $\varepsilon_\perp<0, \varepsilon_\parallel>0)$, the light can be
refracted negatively as a forward
wave.~\cite{poddubny2013hyperbolic,belov2003backward} This refraction
phenomenon can also take place in bulk Rashba system, which can act as
hyperbolic metamaterial at certain frequency
range.~\cite{shibata2016theory} Therefore, due to the forward
transmitted wave, in hyperbolic metamaterial the negative refraction
can take place without having negative effective refractive
index. This situation is different from our case for WSM, where the
negative refraction takes place with backward transmitted wave,
similar to Veselago medium.

If we use $\Omega_{b}=0.5$ we have $\omega_p=800~\textrm{THz}$ and the
corresponding region II can be found within $\left(625\leq\omega\leq
800~\textrm{THz}\right)$. If we use $\omega_p=9~\textrm{THz}$ which is
measured in experiment,~\cite{sushkov2015optical} $\Omega_{b}=44$ and
the corresponding region II can be found within
$\left(0.2\leq\omega\leq 9~\textrm{THz}\right)$.

\begin{figure}[h!]
  \includegraphics[width=85mm]{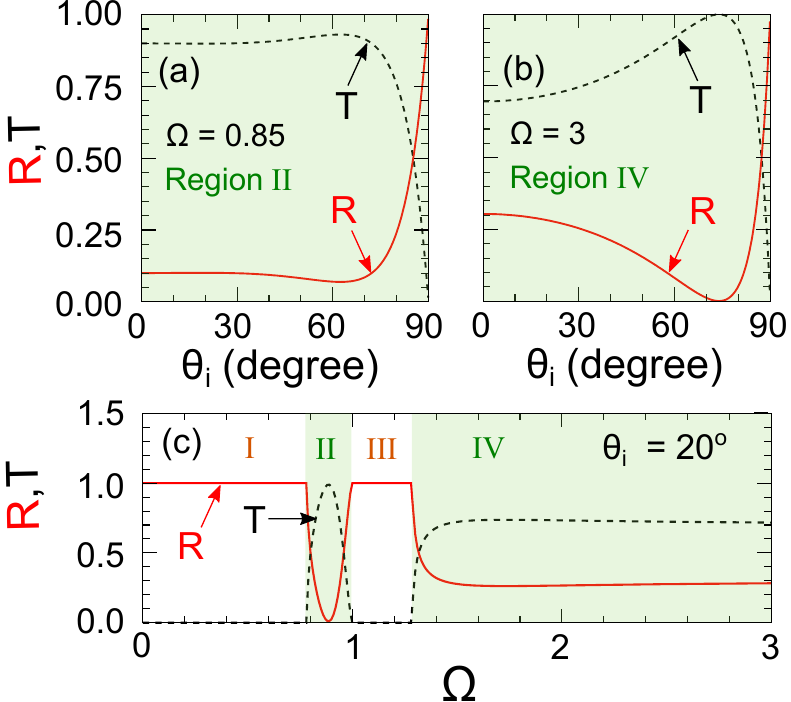}
  \caption{The $R$ and $T$ spectra as a function of $\theta_i$ shown
    as solid and dashed line, respectively, for (a) region II
    ($\Omega=0.85$), (b) region IV ($\Omega=3$). In (a) the negative
    solution of $n_w$ is used, while in (b) the positive one is
    used. In both cases, $R+T=1$. (c) The $R$ and $T$ spectra as a
    function of $\Omega$ with fixed $\theta_i=20^\circ$ shown as solid
    and dashed line, respectively. In shaded regions II and IV, the
    wave is transmitted.  However, only in region II we expect that
    negative refraction could occur. }
\label{fig6}
\end{figure}

Using Eq.~(\ref{eq:intensity}), the transmission probability $T$ is
given by
\begin{align}
  T = \frac{I_t}{I_i} 
  = \frac{1}{n_w^{\pm}}
  \frac{\varepsilon_1}{\varepsilon_0}
  \frac{\cos\theta_t}{\cos\theta_i}\left|t\right|^2,
\label{eq:t}
\end{align}
where $I_i=(c/2)\left| E^{i}_{0}\right|^2\varepsilon_0\cos\theta_i$ is
the incident intensity. The reflection probability $R$ is given by
$R=\left|r\right|^2$. In Figs.~\ref{fig6}(a) and \ref{fig6}(b) we show
the $R$ and $T$ spectra for region II ($\Omega=0.85$) and region IV
($\Omega=3$), where the EM wave propagation is allowed. In the case of
region II, we adopt the $n_w^{-}$, while in the case of region IV, we
adopt $n_w^{+}$. In region IV, the WSM acts as a NM for $\Omega>1$.
Figure~\ref{fig6}(b) shows $R=0$ at $\theta_i=\arctan~n_w$, which
corresponds to the Brewster angle. In both cases, we found $R+T=1$. In
Fig.~\ref{fig6}(c), we plot the $R$ and $T$ spectra as a function of
$\Omega$ at a fixed incident angle $\theta_i=20^\circ$. In region II,
we expect that the negative refraction can take place. In NM, all EM
wave is reflected in the region II due to the imaginary transmitted
wave vector. The region II of WSM, the $R$ gradually decreases with
increasing $\Omega$ because the transmitted wave vector acquires real
value, which signifies the transmission of the incident wave to
WSM. After reaching the minimum of $R$ at $\Omega=0.9$, the reflection
probability increases gradually up to $R=1$ at $\Omega=1$, above which
the transmitted wave vector has only imaginary value that makes
$T=0$. It is important to note that the negative refraction in WSM
occurs only in region II, which has frequency range close to
$\omega_p$, which can be seen in Figs.~\ref{fig2}
and~\ref{fig6}(c). Because $\omega_p$ depends on $E_{\textrm{F}}$ [See
  Eq.~(\ref{eq:wp})], by controlling the $E_{\textrm{F}}$, we can
control the frequency, where negative refraction occurs, which will be
discussed as below.

\begin{figure}[t]
  \includegraphics[width=85mm]{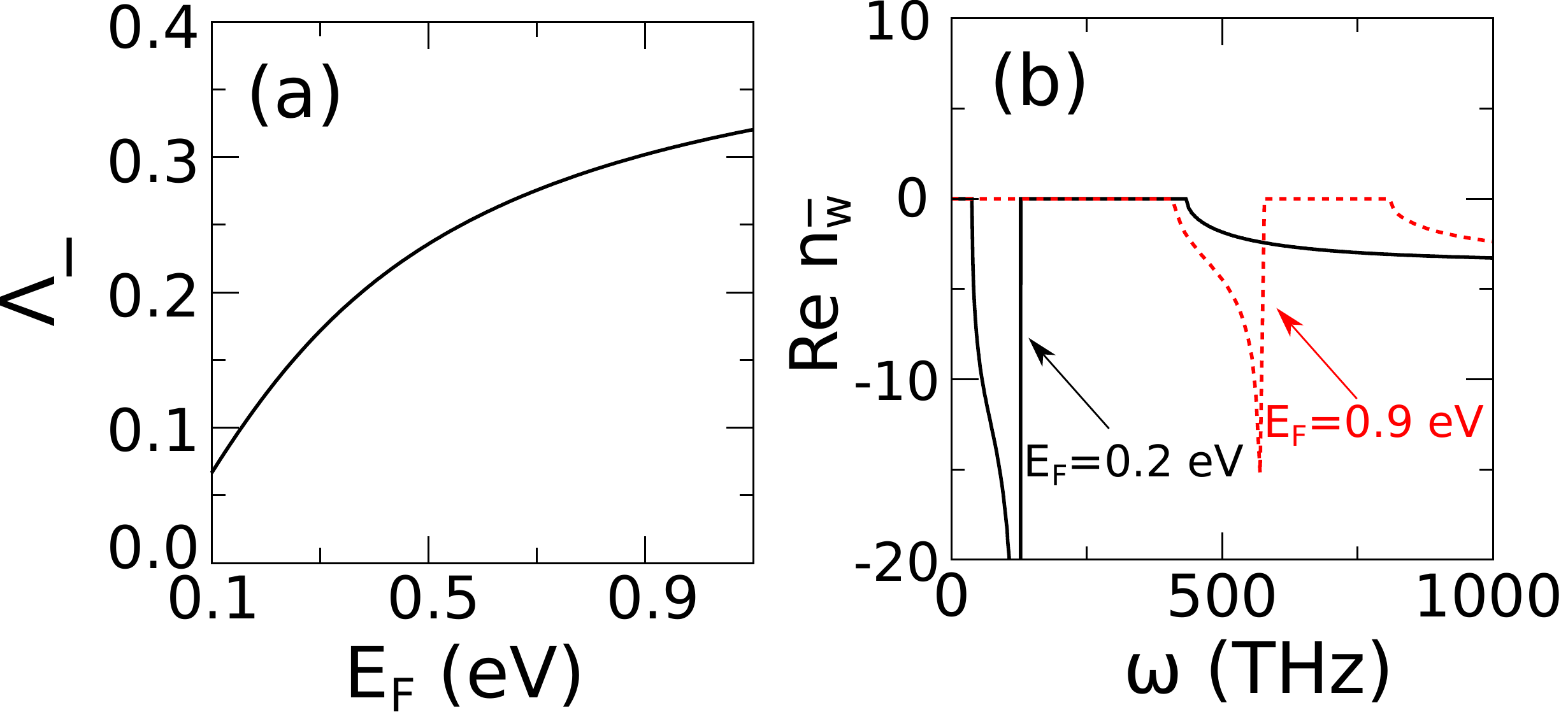}
  \caption{(a) The $\Lambda_-$ as a function of $E_{\textrm{F}}$. (b)
    The real part of $n_w^{-}$ for $E_{\textrm{F}}=0.2~\textrm{eV}$
    and $0.9~\textrm{eV}$.  }
\label{fig7}
\end{figure} 

It is also useful to have a parameter that gives us information
whether or not we have negative refraction for a given
$E_{\textrm{F}}$. By using Eq.~(\ref{eq:wp}), the frequency range of
region II, where we expect the negative refraction, can be rewritten
as
$\Lambda_-<\frac{\hbar\omega}{E_{\textrm{F}}}<\sqrt{\frac{4\alpha}{3\pi}}$,
where $\Lambda_-=\sqrt{\frac{4\alpha}{3\pi}}\Omega_-$ and
$\sqrt{\frac{4\alpha}{3\pi}}=0.423$. $\Omega_-$ is given by
Eq.~(\ref{eq:opm}).  $\Lambda_-$ is plotted in Fig.~\ref{fig7}(a) as a
function of $E_{\textrm{F}}$.  Hence, by taking ratio of EM wave
energy ($\hbar\omega$) and $E_{\textrm{F}}$, we can predict whether
the negative refraction occurs by using Fig.~\ref{fig7}(a). In
Fig.~\ref{fig7}(b), we plot the $n_w^{-}$ as a function of frequency
$(\omega)$ in a real unit for $E_{\textrm{F}}=0.2~\textrm{eV}$ and
$0.9~\textrm{eV}$. Increasing $E_{\textrm{F}}$ will shift the region
II and region IV to higher frequency. The frequency range of region II
monotonically increases with increasing $E_{\textrm{F}}$. Note
that $\alpha$ depends on $v_{\textrm{F}}$ [see Eq.~(\ref{eq:wp})].

\section{Conclusion}
  
In conclusion, we have shown theoretically that negative refraction
can occur in the WSM, which is justified from its reflection spectra.
The refractive index of WSM is negative at a specific frequency range
close to the plasmon frequency. The negative refractive index is
required for the propagation of TM EM wave with frequency smaller than
the plasmon frequency in the direction perpendicular to the separation
of Weyl nodes to conserve the energy and to obtain the physically
correct solution. We suggest that by using only the WSM, it is not
necessary to make a complicated structure of metamaterials to obtain
negative refraction. It would be desired if the phenomenon could be
measured in future experiments.

\begin{acknowledgments}
M.S.U is supported by the MEXT scholarship, Japan.
A.R.T.N. acknowledges the Leading Graduate School in Tohoku
University.  R.S. acknowledges JSPS KAKENHI Grant Numbers JP25107005
and JP25286005.
\end{acknowledgments}

%\bibliographystyle{apsrev4-1} 
%\bibliography{stda2000}

\begin{thebibliography}{34}%
\makeatletter
\providecommand \@ifxundefined [1]{%
 \@ifx{#1\undefined}
}%
\providecommand \@ifnum [1]{%
 \ifnum #1\expandafter \@firstoftwo
 \else \expandafter \@secondoftwo
 \fi
}%
\providecommand \@ifx [1]{%
 \ifx #1\expandafter \@firstoftwo
 \else \expandafter \@secondoftwo
 \fi
}%
\providecommand \natexlab [1]{#1}%
\providecommand \enquote  [1]{``#1''}%
\providecommand \bibnamefont  [1]{#1}%
\providecommand \bibfnamefont [1]{#1}%
\providecommand \citenamefont [1]{#1}%
\providecommand \href@noop [0]{\@secondoftwo}%
\providecommand \href [0]{\begingroup \@sanitize@url \@href}%
\providecommand \@href[1]{\@@startlink{#1}\@@href}%
\providecommand \@@href[1]{\endgroup#1\@@endlink}%
\providecommand \@sanitize@url [0]{\catcode `\\12\catcode `\$12\catcode
  `\&12\catcode `\#12\catcode `\^12\catcode `\_12\catcode `\%12\relax}%
\providecommand \@@startlink[1]{}%
\providecommand \@@endlink[0]{}%
\providecommand \url  [0]{\begingroup\@sanitize@url \@url }%
\providecommand \@url [1]{\endgroup\@href {#1}{\urlprefix }}%
\providecommand \urlprefix  [0]{URL }%
\providecommand \Eprint [0]{\href }%
\providecommand \doibase [0]{http://dx.doi.org/}%
\providecommand \selectlanguage [0]{\@gobble}%
\providecommand \bibinfo  [0]{\@secondoftwo}%
\providecommand \bibfield  [0]{\@secondoftwo}%
\providecommand \translation [1]{[#1]}%
\providecommand \BibitemOpen [0]{}%
\providecommand \bibitemStop [0]{}%
\providecommand \bibitemNoStop [0]{.\EOS\space}%
\providecommand \EOS [0]{\spacefactor3000\relax}%
\providecommand \BibitemShut  [1]{\csname bibitem#1\endcsname}%
\let\auto@bib@innerbib\@empty
%</preamble>
\bibitem [{\citenamefont {Veselago}(1968)}]{veselago1968electrodynamics}%
  \BibitemOpen
  \bibfield  {author} {\bibinfo {author} {\bibfnamefont {V.~G.}\ \bibnamefont
  {Veselago}},\ }\href@noop {} {\bibfield  {journal} {\bibinfo  {journal}
  {Phys. Usp.}\ }\textbf {\bibinfo {volume} {10}},\ \bibinfo {pages} {509}
  (\bibinfo {year} {1968})}\BibitemShut {NoStop}%
\bibitem [{\citenamefont {Cubukcu}\ \emph {et~al.}(2003)\citenamefont
  {Cubukcu}, \citenamefont {Aydin}, \citenamefont {Ozbay}, \citenamefont
  {Foteinopoulou},\ and\ \citenamefont
  {Soukoulis}}]{cubukcu2003electromagnetic}%
  \BibitemOpen
  \bibfield  {author} {\bibinfo {author} {\bibfnamefont {E.}~\bibnamefont
  {Cubukcu}}, \bibinfo {author} {\bibfnamefont {K.}~\bibnamefont {Aydin}},
  \bibinfo {author} {\bibfnamefont {E.}~\bibnamefont {Ozbay}}, \bibinfo
  {author} {\bibfnamefont {S.}~\bibnamefont {Foteinopoulou}}, \ and\ \bibinfo
  {author} {\bibfnamefont {C.~M.}\ \bibnamefont {Soukoulis}},\ }\href@noop {}
  {\bibfield  {journal} {\bibinfo  {journal} {Nature}\ }\textbf {\bibinfo
  {volume} {423}},\ \bibinfo {pages} {604} (\bibinfo {year}
  {2003})}\BibitemShut {NoStop}%
\bibitem [{\citenamefont {Pendry}(2000)}]{pendry2000negative}%
  \BibitemOpen
  \bibfield  {author} {\bibinfo {author} {\bibfnamefont {J.~B.}\ \bibnamefont
  {Pendry}},\ }\href@noop {} {\bibfield  {journal} {\bibinfo  {journal} {Phys.
  Rev. Lett.}\ }\textbf {\bibinfo {volume} {85}},\ \bibinfo {pages} {3966}
  (\bibinfo {year} {2000})}\BibitemShut {NoStop}%
\bibitem [{\citenamefont {Habe}\ and\ \citenamefont
  {Koshino}(2015)}]{habe2015spin}%
  \BibitemOpen
  \bibfield  {author} {\bibinfo {author} {\bibfnamefont {T.}~\bibnamefont
  {Habe}}\ and\ \bibinfo {author} {\bibfnamefont {M.}~\bibnamefont {Koshino}},\
  }\href@noop {} {\bibfield  {journal} {\bibinfo  {journal} {Phys. Rev. B}\
  }\textbf {\bibinfo {volume} {91}},\ \bibinfo {pages} {201407} (\bibinfo
  {year} {2015})}\BibitemShut {NoStop}%
\bibitem [{\citenamefont {Cheianov}\ \emph {et~al.}(2007)\citenamefont
  {Cheianov}, \citenamefont {Fal'ko},\ and\ \citenamefont
  {Altshuler}}]{cheianov2007focusing}%
  \BibitemOpen
  \bibfield  {author} {\bibinfo {author} {\bibfnamefont {V.~V.}\ \bibnamefont
  {Cheianov}}, \bibinfo {author} {\bibfnamefont {V.}~\bibnamefont {Fal'ko}}, \
  and\ \bibinfo {author} {\bibfnamefont {B.}~\bibnamefont {Altshuler}},\
  }\href@noop {} {\bibfield  {journal} {\bibinfo  {journal} {Science}\ }\textbf
  {\bibinfo {volume} {315}},\ \bibinfo {pages} {1252} (\bibinfo {year}
  {2007})}\BibitemShut {NoStop}%
\bibitem [{\citenamefont {Smith}\ \emph {et~al.}(2004)\citenamefont {Smith},
  \citenamefont {Pendry},\ and\ \citenamefont
  {Wiltshire}}]{smith2004metamaterials}%
  \BibitemOpen
  \bibfield  {author} {\bibinfo {author} {\bibfnamefont {D.~R.}\ \bibnamefont
  {Smith}}, \bibinfo {author} {\bibfnamefont {J.~B.}\ \bibnamefont {Pendry}}, \
  and\ \bibinfo {author} {\bibfnamefont {M.~C.~K.}\ \bibnamefont {Wiltshire}},\
  }\href@noop {} {\bibfield  {journal} {\bibinfo  {journal} {Science}\ }\textbf
  {\bibinfo {volume} {305}},\ \bibinfo {pages} {788} (\bibinfo {year}
  {2004})}\BibitemShut {NoStop}%
\bibitem [{\citenamefont {Lu}\ \emph {et~al.}(2003)\citenamefont {Lu},
  \citenamefont {Grzegorczyk}, \citenamefont {Zhang}, \citenamefont
  {Pacheco~Jr}, \citenamefont {Wu}, \citenamefont {Kong},\ and\ \citenamefont
  {Chen}}]{lu2003vcerenkov}%
  \BibitemOpen
  \bibfield  {author} {\bibinfo {author} {\bibfnamefont {J.}~\bibnamefont
  {Lu}}, \bibinfo {author} {\bibfnamefont {T.~M.}\ \bibnamefont {Grzegorczyk}},
  \bibinfo {author} {\bibfnamefont {Y.}~\bibnamefont {Zhang}}, \bibinfo
  {author} {\bibfnamefont {J.}~\bibnamefont {Pacheco~Jr}}, \bibinfo {author}
  {\bibfnamefont {B.-I.}\ \bibnamefont {Wu}}, \bibinfo {author} {\bibfnamefont
  {J.~A.}\ \bibnamefont {Kong}}, \ and\ \bibinfo {author} {\bibfnamefont
  {M.}~\bibnamefont {Chen}},\ }\href@noop {} {\bibfield  {journal} {\bibinfo
  {journal} {Opt. Express}\ }\textbf {\bibinfo {volume} {11}},\ \bibinfo
  {pages} {723} (\bibinfo {year} {2003})}\BibitemShut {NoStop}%
\bibitem [{\citenamefont {Ziemkiewicz}\ and\ \citenamefont
  {Zieli{\'n}ska-Raczy{\'n}ska}(2015)}]{ziemkiewicz2015cherenkov}%
  \BibitemOpen
  \bibfield  {author} {\bibinfo {author} {\bibfnamefont {D.}~\bibnamefont
  {Ziemkiewicz}}\ and\ \bibinfo {author} {\bibfnamefont {S.}~\bibnamefont
  {Zieli{\'n}ska-Raczy{\'n}ska}},\ }\href@noop {} {\bibfield  {journal}
  {\bibinfo  {journal} {J. Opt. Soc. Am. B}\ }\textbf {\bibinfo {volume}
  {32}},\ \bibinfo {pages} {1637} (\bibinfo {year} {2015})}\BibitemShut
  {NoStop}%
\bibitem [{\citenamefont {Shalaev}(2007)}]{shalaev2007optical}%
  \BibitemOpen
  \bibfield  {author} {\bibinfo {author} {\bibfnamefont {V.~M.}\ \bibnamefont
  {Shalaev}},\ }\href@noop {} {\bibfield  {journal} {\bibinfo  {journal} {Nat.
  Photon.}\ }\textbf {\bibinfo {volume} {1}},\ \bibinfo {pages} {41} (\bibinfo
  {year} {2007})}\BibitemShut {NoStop}%
\bibitem [{\citenamefont {Boltasseva}\ and\ \citenamefont
  {Shalaev}(2008)}]{boltasseva2008fabrication}%
  \BibitemOpen
  \bibfield  {author} {\bibinfo {author} {\bibfnamefont {A.}~\bibnamefont
  {Boltasseva}}\ and\ \bibinfo {author} {\bibfnamefont {V.~M.}\ \bibnamefont
  {Shalaev}},\ }\href@noop {} {\bibfield  {journal} {\bibinfo  {journal}
  {Metamaterials}\ }\textbf {\bibinfo {volume} {2}},\ \bibinfo {pages} {1}
  (\bibinfo {year} {2008})}\BibitemShut {NoStop}%
\bibitem [{\citenamefont {Padilla}\ \emph {et~al.}(2006)\citenamefont
  {Padilla}, \citenamefont {Basov},\ and\ \citenamefont
  {Smith}}]{padilla2006negative}%
  \BibitemOpen
  \bibfield  {author} {\bibinfo {author} {\bibfnamefont {W.~J.}\ \bibnamefont
  {Padilla}}, \bibinfo {author} {\bibfnamefont {D.~N.}\ \bibnamefont {Basov}},
  \ and\ \bibinfo {author} {\bibfnamefont {D.~R.}\ \bibnamefont {Smith}},\
  }\href@noop {} {\bibfield  {journal} {\bibinfo  {journal} {Mater. Today}\
  }\textbf {\bibinfo {volume} {9}},\ \bibinfo {pages} {28} (\bibinfo {year}
  {2006})}\BibitemShut {NoStop}%
\bibitem [{\citenamefont {Valentine}\ \emph {et~al.}(2008)\citenamefont
  {Valentine}, \citenamefont {Zhang}, \citenamefont {Zentgraf}, \citenamefont
  {Ulin-Avila}, \citenamefont {Genov}, \citenamefont {Bartal},\ and\
  \citenamefont {Zhang}}]{valentine2008three}%
  \BibitemOpen
  \bibfield  {author} {\bibinfo {author} {\bibfnamefont {J.}~\bibnamefont
  {Valentine}}, \bibinfo {author} {\bibfnamefont {S.}~\bibnamefont {Zhang}},
  \bibinfo {author} {\bibfnamefont {T.}~\bibnamefont {Zentgraf}}, \bibinfo
  {author} {\bibfnamefont {E.}~\bibnamefont {Ulin-Avila}}, \bibinfo {author}
  {\bibfnamefont {D.~A.}\ \bibnamefont {Genov}}, \bibinfo {author}
  {\bibfnamefont {G.}~\bibnamefont {Bartal}}, \ and\ \bibinfo {author}
  {\bibfnamefont {X.}~\bibnamefont {Zhang}},\ }\href@noop {} {\bibfield
  {journal} {\bibinfo  {journal} {Nature}\ }\textbf {\bibinfo {volume} {455}},\
  \bibinfo {pages} {376} (\bibinfo {year} {2008})}\BibitemShut {NoStop}%
\bibitem [{\citenamefont {Ishikawa}\ \emph {et~al.}(2005)\citenamefont
  {Ishikawa}, \citenamefont {Tanaka},\ and\ \citenamefont
  {Kawata}}]{ishikawa2005negative}%
  \BibitemOpen
  \bibfield  {author} {\bibinfo {author} {\bibfnamefont {A.}~\bibnamefont
  {Ishikawa}}, \bibinfo {author} {\bibfnamefont {T.}~\bibnamefont {Tanaka}}, \
  and\ \bibinfo {author} {\bibfnamefont {S.}~\bibnamefont {Kawata}},\
  }\href@noop {} {\bibfield  {journal} {\bibinfo  {journal} {Phys. Rev. Lett.}\
  }\textbf {\bibinfo {volume} {95}},\ \bibinfo {pages} {237401} (\bibinfo
  {year} {2005})}\BibitemShut {NoStop}%
\bibitem [{\citenamefont {Moser}\ \emph {et~al.}(2005)\citenamefont {Moser},
  \citenamefont {Casse}, \citenamefont {Wilhelmi},\ and\ \citenamefont
  {Saw}}]{moser2005terahertz}%
  \BibitemOpen
  \bibfield  {author} {\bibinfo {author} {\bibfnamefont {H.}~\bibnamefont
  {Moser}}, \bibinfo {author} {\bibfnamefont {B.}~\bibnamefont {Casse}},
  \bibinfo {author} {\bibfnamefont {O.}~\bibnamefont {Wilhelmi}}, \ and\
  \bibinfo {author} {\bibfnamefont {B.}~\bibnamefont {Saw}},\ }\href@noop {}
  {\bibfield  {journal} {\bibinfo  {journal} {Phys. Rev. Lett.}\ }\textbf
  {\bibinfo {volume} {94}},\ \bibinfo {pages} {063901} (\bibinfo {year}
  {2005})}\BibitemShut {NoStop}%
\bibitem [{\citenamefont {Bilotti}\ \emph {et~al.}(2007)\citenamefont
  {Bilotti}, \citenamefont {Toscano},\ and\ \citenamefont
  {Vegni}}]{bilotti2007design}%
  \BibitemOpen
  \bibfield  {author} {\bibinfo {author} {\bibfnamefont {F.}~\bibnamefont
  {Bilotti}}, \bibinfo {author} {\bibfnamefont {A.}~\bibnamefont {Toscano}}, \
  and\ \bibinfo {author} {\bibfnamefont {L.}~\bibnamefont {Vegni}},\
  }\href@noop {} {\bibfield  {journal} {\bibinfo  {journal} {IEEE Trans.
  Antennas. Propag.}\ }\textbf {\bibinfo {volume} {55}},\ \bibinfo {pages}
  {2258} (\bibinfo {year} {2007})}\BibitemShut {NoStop}%
\bibitem [{\citenamefont {Parimi}\ \emph {et~al.}(2003)\citenamefont {Parimi},
  \citenamefont {Lu}, \citenamefont {Vodo},\ and\ \citenamefont
  {Sridhar}}]{parimi2003photonic}%
  \BibitemOpen
  \bibfield  {author} {\bibinfo {author} {\bibfnamefont {P.~V.}\ \bibnamefont
  {Parimi}}, \bibinfo {author} {\bibfnamefont {W.~T.}\ \bibnamefont {Lu}},
  \bibinfo {author} {\bibfnamefont {P.}~\bibnamefont {Vodo}}, \ and\ \bibinfo
  {author} {\bibfnamefont {S.}~\bibnamefont {Sridhar}},\ }\href@noop {}
  {\bibfield  {journal} {\bibinfo  {journal} {Nature}\ }\textbf {\bibinfo
  {volume} {426}},\ \bibinfo {pages} {404} (\bibinfo {year}
  {2003})}\BibitemShut {NoStop}%
\bibitem [{\citenamefont {Hofmann}\ and\ \citenamefont
  {Das~Sarma}(2016)}]{hofmann2016surface}%
  \BibitemOpen
  \bibfield  {author} {\bibinfo {author} {\bibfnamefont {J.}~\bibnamefont
  {Hofmann}}\ and\ \bibinfo {author} {\bibfnamefont {S.}~\bibnamefont
  {Das~Sarma}},\ }\href@noop {} {\bibfield  {journal} {\bibinfo  {journal}
  {Phys. Rev. B}\ }\textbf {\bibinfo {volume} {93}},\ \bibinfo {pages} {241402}
  (\bibinfo {year} {2016})}\BibitemShut {NoStop}%
\bibitem [{\citenamefont {Burkov}\ and\ \citenamefont
  {Balents}(2011)}]{burkov2011weyl}%
  \BibitemOpen
  \bibfield  {author} {\bibinfo {author} {\bibfnamefont {A.~A.}\ \bibnamefont
  {Burkov}}\ and\ \bibinfo {author} {\bibfnamefont {L.}~\bibnamefont
  {Balents}},\ }\href@noop {} {\bibfield  {journal} {\bibinfo  {journal} {Phys.
  Rev. Lett.}\ }\textbf {\bibinfo {volume} {107}},\ \bibinfo {pages} {127205}
  (\bibinfo {year} {2011})}\BibitemShut {NoStop}%
\bibitem [{\citenamefont {Vazifeh}\ and\ \citenamefont
  {Franz}(2013)}]{vazifeh2013electromagnetic}%
  \BibitemOpen
  \bibfield  {author} {\bibinfo {author} {\bibfnamefont {M.~M.}\ \bibnamefont
  {Vazifeh}}\ and\ \bibinfo {author} {\bibfnamefont {M.}~\bibnamefont
  {Franz}},\ }\href@noop {} {\bibfield  {journal} {\bibinfo  {journal} {Phys.
  Rev. Lett.}\ }\textbf {\bibinfo {volume} {111}},\ \bibinfo {pages} {027201}
  (\bibinfo {year} {2013})}\BibitemShut {NoStop}%
\bibitem [{\citenamefont {Koshino}\ and\ \citenamefont
  {Hizbullah}(2016)}]{koshino2016magnetic}%
  \BibitemOpen
  \bibfield  {author} {\bibinfo {author} {\bibfnamefont {M.}~\bibnamefont
  {Koshino}}\ and\ \bibinfo {author} {\bibfnamefont {I.~F.}\ \bibnamefont
  {Hizbullah}},\ }\href@noop {} {\bibfield  {journal} {\bibinfo  {journal}
  {Phys. Rev. B}\ }\textbf {\bibinfo {volume} {93}},\ \bibinfo {pages} {045201}
  (\bibinfo {year} {2016})}\BibitemShut {NoStop}%
\bibitem [{\citenamefont {Ominato}\ and\ \citenamefont
  {Koshino}(2015)}]{ominato2015quantum}%
  \BibitemOpen
  \bibfield  {author} {\bibinfo {author} {\bibfnamefont {Y.}~\bibnamefont
  {Ominato}}\ and\ \bibinfo {author} {\bibfnamefont {M.}~\bibnamefont
  {Koshino}},\ }\href@noop {} {\bibfield  {journal} {\bibinfo  {journal} {Phys.
  Rev. B}\ }\textbf {\bibinfo {volume} {91}},\ \bibinfo {pages} {035202}
  (\bibinfo {year} {2015})}\BibitemShut {NoStop}%
\bibitem [{\citenamefont {Sushkov}\ \emph {et~al.}(2015)\citenamefont
  {Sushkov}, \citenamefont {Hofmann}, \citenamefont {Jenkins}, \citenamefont
  {Ishikawa}, \citenamefont {Nakatsuji}, \citenamefont {Das~Sarma},\ and\
  \citenamefont {Drew}}]{sushkov2015optical}%
  \BibitemOpen
  \bibfield  {author} {\bibinfo {author} {\bibfnamefont {A.~B.}\ \bibnamefont
  {Sushkov}}, \bibinfo {author} {\bibfnamefont {J.~B.}\ \bibnamefont
  {Hofmann}}, \bibinfo {author} {\bibfnamefont {G.~S.}\ \bibnamefont
  {Jenkins}}, \bibinfo {author} {\bibfnamefont {J.}~\bibnamefont {Ishikawa}},
  \bibinfo {author} {\bibfnamefont {S.}~\bibnamefont {Nakatsuji}}, \bibinfo
  {author} {\bibfnamefont {S.}~\bibnamefont {Das~Sarma}}, \ and\ \bibinfo
  {author} {\bibfnamefont {H.~D.}\ \bibnamefont {Drew}},\ }\href@noop {}
  {\bibfield  {journal} {\bibinfo  {journal} {Phys. Rev. B}\ }\textbf {\bibinfo
  {volume} {92}},\ \bibinfo {pages} {241108} (\bibinfo {year}
  {2015})}\BibitemShut {NoStop}%
\bibitem [{\citenamefont {Zyuzin}\ and\ \citenamefont
  {Zyuzin}(2015)}]{zyuzin2015chiral}%
  \BibitemOpen
  \bibfield  {author} {\bibinfo {author} {\bibfnamefont {A.~A.}\ \bibnamefont
  {Zyuzin}}\ and\ \bibinfo {author} {\bibfnamefont {V.~A.}\ \bibnamefont
  {Zyuzin}},\ }\href@noop {} {\bibfield  {journal} {\bibinfo  {journal} {Phys.
  Rev. B}\ }\textbf {\bibinfo {volume} {92}},\ \bibinfo {pages} {115310}
  (\bibinfo {year} {2015})}\BibitemShut {NoStop}%
\bibitem [{\citenamefont {Grushin}(2012)}]{grushin2012consequences}%
  \BibitemOpen
  \bibfield  {author} {\bibinfo {author} {\bibfnamefont {A.~G.}\ \bibnamefont
  {Grushin}},\ }\href@noop {} {\bibfield  {journal} {\bibinfo  {journal} {Phys.
  Rev. D}\ }\textbf {\bibinfo {volume} {86}},\ \bibinfo {pages} {045001}
  (\bibinfo {year} {2012})}\BibitemShut {NoStop}%
\bibitem [{\citenamefont {Zyuzin}\ and\ \citenamefont
  {Burkov}(2012)}]{zyuzin2012topological}%
  \BibitemOpen
  \bibfield  {author} {\bibinfo {author} {\bibfnamefont {A.~A.}\ \bibnamefont
  {Zyuzin}}\ and\ \bibinfo {author} {\bibfnamefont {A.~A.}\ \bibnamefont
  {Burkov}},\ }\href@noop {} {\bibfield  {journal} {\bibinfo  {journal} {Phys.
  Rev. B}\ }\textbf {\bibinfo {volume} {86}},\ \bibinfo {pages} {115133}
  (\bibinfo {year} {2012})}\BibitemShut {NoStop}%
\bibitem [{\citenamefont {Zyuzin}\ \emph {et~al.}(2012)\citenamefont {Zyuzin},
  \citenamefont {Wu},\ and\ \citenamefont {Burkov}}]{zyuzin2012weyl}%
  \BibitemOpen
  \bibfield  {author} {\bibinfo {author} {\bibfnamefont {A.~A.}\ \bibnamefont
  {Zyuzin}}, \bibinfo {author} {\bibfnamefont {S.}~\bibnamefont {Wu}}, \ and\
  \bibinfo {author} {\bibfnamefont {A.~A.}\ \bibnamefont {Burkov}},\
  }\href@noop {} {\bibfield  {journal} {\bibinfo  {journal} {Phys. Rev. B}\
  }\textbf {\bibinfo {volume} {85}},\ \bibinfo {pages} {165110} (\bibinfo
  {year} {2012})}\BibitemShut {NoStop}%
\bibitem [{\citenamefont {Hosur}\ and\ \citenamefont
  {Qi}(2013)}]{hosur2013recent}%
  \BibitemOpen
  \bibfield  {author} {\bibinfo {author} {\bibfnamefont {P.}~\bibnamefont
  {Hosur}}\ and\ \bibinfo {author} {\bibfnamefont {X.}~\bibnamefont {Qi}},\
  }\href@noop {} {\bibfield  {journal} {\bibinfo  {journal} {C. R. Phys.}\
  }\textbf {\bibinfo {volume} {14}},\ \bibinfo {pages} {857} (\bibinfo {year}
  {2013})}\BibitemShut {NoStop}%
\bibitem [{\citenamefont {Saleh}\ and\ \citenamefont
  {Teich}(1991)}]{saleh1991fundamentals}%
  \BibitemOpen
  \bibfield  {author} {\bibinfo {author} {\bibfnamefont {B.~E.~A.}\
  \bibnamefont {Saleh}}\ and\ \bibinfo {author} {\bibfnamefont {M.~C.}\
  \bibnamefont {Teich}},\ }\href@noop {} {\emph {\bibinfo {title} {Fundamentals
  of photonics}}}\ (\bibinfo  {publisher} {Wiley, New York},\ \bibinfo {year}
  {1991})\BibitemShut {NoStop}%
\bibitem [{\citenamefont {Smith}\ and\ \citenamefont
  {Kroll}(2000)}]{smith2000negative}%
  \BibitemOpen
  \bibfield  {author} {\bibinfo {author} {\bibfnamefont {D.~R.}\ \bibnamefont
  {Smith}}\ and\ \bibinfo {author} {\bibfnamefont {N.}~\bibnamefont {Kroll}},\
  }\href@noop {} {\bibfield  {journal} {\bibinfo  {journal} {Phys. Rev. Lett.}\
  }\textbf {\bibinfo {volume} {85}},\ \bibinfo {pages} {2933} (\bibinfo {year}
  {2000})}\BibitemShut {NoStop}%
\bibitem [{\citenamefont {Ramakrishna}(2005)}]{ramakrishna2005physics}%
  \BibitemOpen
  \bibfield  {author} {\bibinfo {author} {\bibfnamefont {S.~A.}\ \bibnamefont
  {Ramakrishna}},\ }\href@noop {} {\bibfield  {journal} {\bibinfo  {journal}
  {Rep. Prog. Phys.}\ }\textbf {\bibinfo {volume} {68}},\ \bibinfo {pages}
  {449} (\bibinfo {year} {2005})}\BibitemShut {NoStop}%
\bibitem [{\citenamefont {Woodley}\ and\ \citenamefont
  {Mojahedi}(2006)}]{woodley2006backward}%
  \BibitemOpen
  \bibfield  {author} {\bibinfo {author} {\bibfnamefont {J.}~\bibnamefont
  {Woodley}}\ and\ \bibinfo {author} {\bibfnamefont {M.}~\bibnamefont
  {Mojahedi}},\ }\href@noop {} {\bibfield  {journal} {\bibinfo  {journal} {J.
  Opt. Soc. Am. B}\ }\textbf {\bibinfo {volume} {23}},\ \bibinfo {pages} {2377}
  (\bibinfo {year} {2006})}\BibitemShut {NoStop}%
\bibitem [{\citenamefont {Poddubny}\ \emph {et~al.}(2013)\citenamefont
  {Poddubny}, \citenamefont {Iorsh}, \citenamefont {Belov},\ and\ \citenamefont
  {Kivshar}}]{poddubny2013hyperbolic}%
  \BibitemOpen
  \bibfield  {author} {\bibinfo {author} {\bibfnamefont {A.}~\bibnamefont
  {Poddubny}}, \bibinfo {author} {\bibfnamefont {I.}~\bibnamefont {Iorsh}},
  \bibinfo {author} {\bibfnamefont {P.}~\bibnamefont {Belov}}, \ and\ \bibinfo
  {author} {\bibfnamefont {Y.}~\bibnamefont {Kivshar}},\ }\href@noop {}
  {\bibfield  {journal} {\bibinfo  {journal} {Nat. Photon.}\ }\textbf {\bibinfo
  {volume} {7}},\ \bibinfo {pages} {948} (\bibinfo {year} {2013})}\BibitemShut
  {NoStop}%
\bibitem [{\citenamefont {Belov}(2003)}]{belov2003backward}%
  \BibitemOpen
  \bibfield  {author} {\bibinfo {author} {\bibfnamefont {P.~A.}\ \bibnamefont
  {Belov}},\ }\href@noop {} {\bibfield  {journal} {\bibinfo  {journal} {Microw.
  Opt. Technol. Lett.}\ }\textbf {\bibinfo {volume} {37}},\ \bibinfo {pages}
  {259} (\bibinfo {year} {2003})}\BibitemShut {NoStop}%
\bibitem [{\citenamefont {Shibata}\ \emph {et~al.}(2016)\citenamefont
  {Shibata}, \citenamefont {Takeuchi}, \citenamefont {Kohno},\ and\
  \citenamefont {Tatara}}]{shibata2016theory}%
  \BibitemOpen
  \bibfield  {author} {\bibinfo {author} {\bibfnamefont {J.}~\bibnamefont
  {Shibata}}, \bibinfo {author} {\bibfnamefont {A.}~\bibnamefont {Takeuchi}},
  \bibinfo {author} {\bibfnamefont {H.}~\bibnamefont {Kohno}}, \ and\ \bibinfo
  {author} {\bibfnamefont {G.}~\bibnamefont {Tatara}},\ }\href@noop {}
  {\bibfield  {journal} {\bibinfo  {journal} {J. Phys. Soc. Japan}\ }\textbf
  {\bibinfo {volume} {85}},\ \bibinfo {pages} {033701} (\bibinfo {year}
  {2016})}\BibitemShut {NoStop}%
\end{thebibliography}


\end{document}